\documentclass[preprint,12pt]{elsarticle}

\usepackage{amssymb}

\newcounter{bla}

\usepackage{listings}
\usepackage{color}
\definecolor{mygray}{rgb}{0.5,0.5,0.5}
\newcommand{\ve}[1]{\mathbf{#1}}

\journal{Computer Physics Communications}

\begin{document}

\begin{frontmatter}



\title{kMap.py: A Python program for simulation and data analysis in photoemission tomography}


\author[a]{Dominik Brandstetter}
\author[a,b,c,d]{Xiaosheng Yang}
\author[a]{Daniel L\"uftner}
\author[b,c,d]{F. Stefan Tautz}
\author[a]{Peter Puschnig\corref{author}}

\cortext[author] {Corresponding author.\\\textit{E-mail address:} peter.puschnig@uni-graz.at}
\address[a]{Karl-Franzens-Universit\"at Graz, Institut f\"ur Physik, NAWI Graz, 8010 Graz, Austria}
\address[b]{Peter Gr\"unberg Institut (PGI-3), Forschungszentrum J\"ulich, 52425 J\"ulich, Germany}
\address[c]{J\"ulich Aachen Research Alliance (JARA), Fundamentals of Future Information Technology, 52425 J\"ulich, Germany}
\address[d]{Experimentalphysik IV A, RWTH Aachen University, 52074 Aachen, Germany}

\begin{abstract}
Ultra-violet photoemission spectroscopy is a widely-used experimental technique to investigate the valence electronic structure of surfaces and interfaces. When detecting the intensity of the emitted electrons not only as a function of their kinetic energy, but also depending on their emission angle, as is done in angle-resolved photoemission spectroscopy (ARPES), extremely rich information about the electronic structure of the investigated sample can be extracted. For organic molecules adsorbed as well-oriented ultra-thin films on metallic surfaces, ARPES has evolved into a technique called photoemission tomography (PT). By approximating the final state of the photoemitted electron as a free electron, PT uses the angular dependence of the photocurrent, a so-called momentum map or $k$-map, and interprets it as the Fourier transform of the initial state's molecular orbital, thereby gaining insights into the geometric and electronic structure of organic/metal interfaces.

In this contribution, we present \verb|kMap.py| which is a Python program that enables the user, via a \verb|PyQt|-based graphical user interface, to simulate photoemission momentum maps of molecular orbitals and to perform a one-to-one comparison between simulation and experiment. Based on the plane wave approximation for the final state, simulated momentum maps are computed numerically from a fast Fourier transform (FFT) of real space molecular orbital distributions, which are used as program input and taken from density functional calculations. The program allows the user to vary a number of simulation parameters, such as the final state kinetic energy, the molecular orientation or the polarization state of the incident light field. Moreover, also experimental photoemission data can be loaded into the program, enabling a direct visual comparison as well as an automatic optimization procedure to determine structural parameters of the molecules or weights of molecular orbitals contributions. With an increasing number of experimental groups employing photoemission tomography to study molecular adsorbate layers, we expect \verb|kMap.py| to serve as a helpful  analysis software to further extend the applicability of PT.

%
%
%
%

\end{abstract}


\end{frontmatter}



{\bf PROGRAM SUMMARY}

\begin{small}
\noindent
{\em Program Title:}  \verb|kMap.py|                                        \\
{\em CPC Library link to program files:} (to be added by Technical Editor) \\
{\em Developer's respository link:}  https://github.com/brands-d/kMap/ \\
{\em Code Ocean capsule:} (to be added by Technical Editor)\\
{\em Licensing provisions:} GPLv3  \\
{\em Programming language:} Python 3.x                                  \\
{\em Nature of problem:}\\
  Photoemission tomography (PT) has evolved as a powerful experimental method to investigate the electronic and geometric structure of organic molecular films [1]. It is based on valence band angle-resolved photoemission spectroscopy and seeks an interpretation of the angular dependence of the photocurrent, a so-called momentum map, from a given initial state in terms of the spatial structure of molecular orbitals. For this purpose, PT heavily relies on a simulation platform which is capable of efficiently predicting momentum maps for a variety of organic molecules, which allows for a convenient way of treating the effect of molecular orientations, and which also accounts for other experimental parameters such as the geometrical setup and nature of the incident photon source. Thereby, PT has been used to determine molecular geometries, gain insight into the nature of the surface chemistry, unambiguously determine the orbital energy ordering in molecular homo- and heterostructures and even reconstruct the orbitals of adsorbed molecules [1--4].
\\
{\em Solution method:}\\
\verb|kMap.py| is a Python program that enables the user, via a \verb|PyQt|-based graphical user interface, to simulate photoemission momentum maps of molecular orbitals and to perform a one-to-one comparison between simulation and experiment. Based on the plane wave approximation for the final state, simulated momentum maps are computed numerically from a fast Fourier transform (FFT) of real space molecular orbital distributions [2] which are used as program input and which are usually obtained from density functional calculations. The user can vary a number of simulation parameters such as the final state kinetic energy, the molecular orientation or the polarization state of the incident light field. Moreover, also experimental photoemission data can be loaded into the program, enabling a direct visual comparison as well as an automatic optimization procedure to minimize the difference between simulated and measured momentum maps. Thereby, structural parameters of the molecules [2] and the weights of molecular orbitals to experimentally observed emission features can be determined [3].
   \\

\end{small}

\section{Introduction}
\label{sec:intro}

In the last decade, photoemission tomography (PT) has evolved as a powerful technique in surface science to analyze the spatial structure of electron orbitals of organic molecules by utilizing data from angle-resolved photoemission spectroscopy (ARPES) experiments \cite{Woodruff2016,Puschnig2017}. When approximating the final state of the photoemitted electron as a plane wave, it has been demonstrated \cite{Gadzuk1974,Puschnig2009a} that the angular distribution of the photocurrent is related to the Fourier transform of the initial molecular orbital. As a combined experimental/theoretical approach, PT seeks an interpretation of the photoelectron angular distribution over a wide angular range, so-called momentum maps, in terms of the molecular orbital structure of the initial state as computed with density functional theory. 
Although the underlying simplification has led to some controversy in the community \cite{Bradshaw2015,Egger2018}, PT has found many interesting applications. These include the unambiguous assignment of emissions to molecular orbital densities  \cite{Puschnig2009a,Wiessner2012,Liu2014,Zamborlini2017,Kliuiev2019}, the deconvolution of spectra into individual orbital contributions beyond the limits of energy resolution \cite{Puschnig2011,Dauth2011} or the extraction of detailed
geometric information \cite{Feyer2014,Park2016,Schonauer2016,Puschnig2017,Kliuiev2019}. 
 
On the experimental side, there has been significant progress in photoemission spectrometers and excitation sources. This includes spin-sensitive detectors, photoemission momentum microscopes, or time-resolved photoelectron spectrometers to be combined with laser excitations sources for pump-probe experiments \cite{Wallauer2020}. With the increasing number of experimental groups employing photoemission tomography to study adsorbate layers, the need for an appropriate analysis software has grown. With \verb|kMap.py| we provide a Python program  that enables the user, via a \verb|PyQt|-based graphical user interface (GUI), to simulate photoemission momentum maps and to perform a one-to-one comparison between simulation and experiment. Moreover, it also allows the user to vary a number of parameters in the simulated momentum maps, such as molecular orientation and/or weights of orbitals, and to run an automatic optimization procedure to minimize the difference between simulation and experiment, a procedure known as \emph{deconvolution} \cite{Puschnig2011}. Compared to other recently announced programs or computational schemes for simulating  angle-resolved photoemission experiments \cite{Moser2016a,Feidt2019,Day2019}, the focus of \verb|kMap.py| lies on molecular systems rather than extended two-dimensional materials. 
 
The paper is organized as follows. In Sec.~\ref{sec:theory}, we briefly review the theoretical background of photoemission tomography, before in Sec.~\ref{sec:method} we discuss the main computational methods as implemented in \verb|kMap.py|. Section~\ref{sec:applications} presents the results of a few benchmark applications of \verb|kMap.py|, and finally Sec.~\ref{sec:outlook} discusses possible future directions for the further development of photoemission tomography.

\section{Theoretical Background}
\label{sec:theory}

\subsection{Photoemission tomography}

Two assumptions lie at the heart of the photoemission tomography technique allowing one to interpret the angular distribution of the photoemitted electrons as Fourier transforms of the initial state orbital. The first states that the initial state is considered to be a single-electron state. While such a molecular orbital interpretation has proved invaluable for understanding many-particle systems and is frequently and successfully used in particular in the field of organic semiconductors, from a quantum mechanical, many-body point of view,  ``a single electron'' in a many-electron system is an object whose observability may be questioned on fundamental grounds \cite{Truhlar2019,Krylov2020}. Indeed, the fundamental quantity to be observed turns out to be the Dyson orbital rather than a molecular orbital. However, in many circumstances these two  different orbital concepts can be expected to lead to essentially identical momentum space orbital densities \cite{Dauth2014}.

\begin{figure}[tb]
\begin{center}
\includegraphics[width=0.75\textwidth]{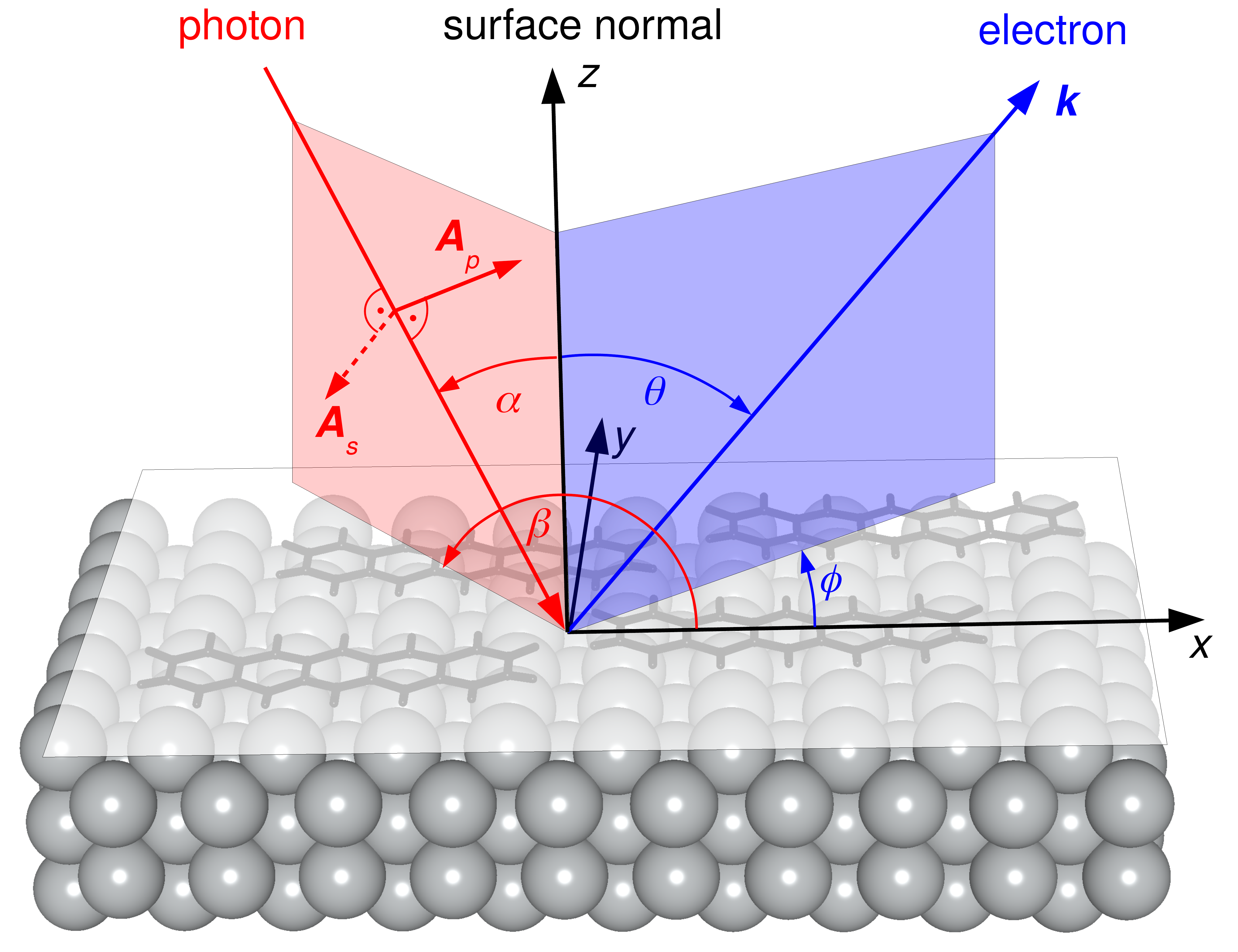}
\end{center}
\caption[Angle-resolved photoemission]
{Definition of the experimental geometry with respect to the Cartesian coordinate frame $(x,y,z)$, where $x$ and $y$ are parallel to the surface and $z$ is along the surface normal. The plane of the incident photon beam (red) is defined by the polar angle $\alpha$ and the azimuthal angle $\beta$. The two principal polarization directions, in-plane ($\ve{A}_p$) and out-of-plane ($\ve{A}_s$), respectively, are indicated. The emitted electron (blue) is characterized by the wave vector $\ve{k}$ and the polar and azimuthal angles $\theta$ and $\phi$ leading to the parallel wave vector components given in Eqs.~\ref{eq:kx} and \ref{eq:ky}, respectively.}
\label{fig:arpes}
\end{figure}

A theoretical description of the angle-resolved photoelectron intensity is generally rather involved, and attempts  to compute it in a quantitative manner are rather scarce. Within this work, photo-excitation is treated as a single coherent process from a molecular orbital to the final state, which is referred to as the one-step-model of photoemission (PE). The PE intensity $I(k_x,k_y;E_\mathrm{kin})$ is given by Fermi's golden rule \cite{Feibelman1974}
\begin{equation}
\label{eq:Feibelman}
I(k_x,k_y;E_\mathrm{kin}) \propto \sum_{i}
                 \left| \langle \Psi_f(k_x,k_y;E_\mathrm{kin}) |
                 \ve{A} \cdot \ve{p} | \Psi_i \rangle \right|^2
                 \times \delta \left(E_i + \Phi + E_\mathrm{kin} - \hbar \omega \right).
\end{equation}
Here, $k_x$ and $k_y$ are the components of the emitted electron's wave vector parallel to the surface, which are related to the polar and azimuthal emission angles $\theta$ and $\phi$ defined in Fig.~\ref{fig:arpes} as follows,
\begin{eqnarray}
k_x & = & k \sin \theta \cos \phi  \label{eq:kx} \\
k_y & = & k \sin \theta \sin \phi, \label{eq:ky}
\end{eqnarray}
where $k$ is the wave number of the emitted electron, thus $E_\mathrm{kin} = \frac{\hbar^2 k^2}{2m}$, where $\hbar$ is the reduced Planck constant and $m$ is the electron mass.

The photocurrent of Eq.~\ref{eq:Feibelman}  is given by a sum over all transitions from occupied initial states $i$ described by wave functions $\Psi_i$ to the final state $\Psi_f$ characterized by the direction $(\theta,\phi)$ and the kinetic energy of the emitted electron. The delta function ensures energy conservation, where $\Phi$ denotes the sample work function, $E_i$ the binding energy of the initial state, and $\hbar \omega$ the energy of the exciting photon. The transition matrix element is given in the dipole approximation, where $\ve{p}$ and $\ve{A}$, respectively,  denote the momentum operator of the electron and the vector potential of the exciting electromagnetic wave.

The difficult part in evaluating Eq.~(\ref{eq:Feibelman}) is the proper treatment of the final state. Here, the second assumption  of photoemission tomography comes into play. In the simplest approach considered here, the final state is approximated by a plane wave (PW), which is characterized only by the direction and wave number of the emitted electron. This has already been proposed more than 30 years  ago \cite{Gadzuk1974} with some success in explaining the observed PE distribution from atoms and small molecules adsorbed at surfaces. Using a plane-wave approximation is appealing since the evaluation of Eq.~(\ref{eq:Feibelman}) renders the photocurrent $I_i$ arising from one particular initial state $i$ proportional to the Fourier transform  $\tilde{\Psi}_{i} (\ve{k})$ of the initial state wave function corrected by the polarization factor $\ve{A} \cdot \ve{k}$:
\begin{equation}
\label{eq:PE1}
I_i(k_x,k_y)  \propto \left|\ve{A} \cdot \ve{k}\right|^2  \cdot \left| \tilde{\Psi}_{i} (\ve{k}) \right|^2. 
\end{equation}
Thus, if the angle-dependent photocurrent of individual initial states can be selectively measured 
(as it can for organic molecules where the intermolecular band dispersion is often smaller than the
energetic separation of individual orbitals), a one-to-one relation between the photocurrent and
the molecular orbitals in reciprocal space can be established.
This allows the measurement of the absolute value of the initial state wave function in reciprocal
space. Note that in some cases, even a reconstruction of molecular orbital densities in real space via a subsequent Fourier transform and an iterative phase retrieval algorithm has been demonstrated \cite{Luftner2013,Weiss2015,Puschnig2017,Graus2019}.

Regarding the applicabilty of the plane wave final state approximation, it has been argued \cite{Goldberg1978,Puschnig2009a} that it can be expected to be valid if the following conditions are fulfilled: (i) $\pi$ orbital emissions from large planar molecules, (ii)
an experimental geometry in which the angle between the polarization vector $\ve{A}$ and the direction of 
the emitted electron $\ve{k}$ is rather small, and (iii) molecules consisting of
many light atoms (H, C, N, O). The latter requirement is a result of the small scattering cross 
section of light atoms  and the presence of many scattering centers is expected to lead to a rather 
weak and structureless angular pattern \cite{Shirley1995,Kera2006}. With these conditions 
satisfied, a one-to-one mapping between the PE intensity and individual molecular orbitals in 
reciprocal space is possible.

\subsection{Initial state}
\label{sec:initialstate}

The evaluation of Eq.~\ref{eq:Feibelman} demands the knowledge of the real space distribution of $\Psi_i(x,y,z)$ of the initial state. As mentioned above, a common approach is to approximate the initial molecular orbitals by the Kohn-Sham states which are obtained from a self-consistent calculation within the framework of density functional theory (DFT). Thus,  the $\Psi_i(x,y,z)$ are given as the eigenfunctions of effective single-particle Schr\"odinger equations, \emph{i.e.}, the Kohn-Sham equations \cite{Kohn1965a}, which, in atomic units, are given by:
\begin{equation}
\left[- \frac{1}{2} \Delta + V_\mathrm{eff}(x,y,z)  \right] \Psi_i(x,y,z) = E_i \Psi_i(x,y,z)
\end{equation}
Here, $V_\mathrm{eff}(x,y,z)$ is the Kohn-Sham potential comprising the external potential due to the atomic nuclei, the Hartree potential and the exchange correlation potential. More details about DFT, which is outside the scope of this contribution, can for instance be found in Refs.~\cite{Fiolhais2003,Sholl2009}. In the context of \verb|kMap.py|, the Kohn-Sham orbitals are orbitals of a finite molecule or cluster of atoms, thus any DFT implementation suitable for such a non-periodic boundary situation can be applied. As explained in Sec.~\ref{sec:database}, the only additional requirement is that the DFT code must be able to write out the Kohn-Sham orbital on a regular three-dimensional grid.

\section{Computational details}
\label{sec:method}

\subsection{Computation of momentum maps}

As can be seen from Eq.~\ref{eq:PE1}, the computation of the intensity of the photoemission angular distribution, the momentum map $I_i(k_x,k_y)$, consists of two terms, namely (i) the Fourier transform of the molecular orbital $\tilde{\Psi}_{i} (\ve{k})$ and (ii) the polarization factor $\left|\ve{A} \cdot \ve{k}\right|^2 $. 

\subsubsection{Discrete Fourier transform}

\begin{figure}[tb]
\includegraphics[width=\textwidth]{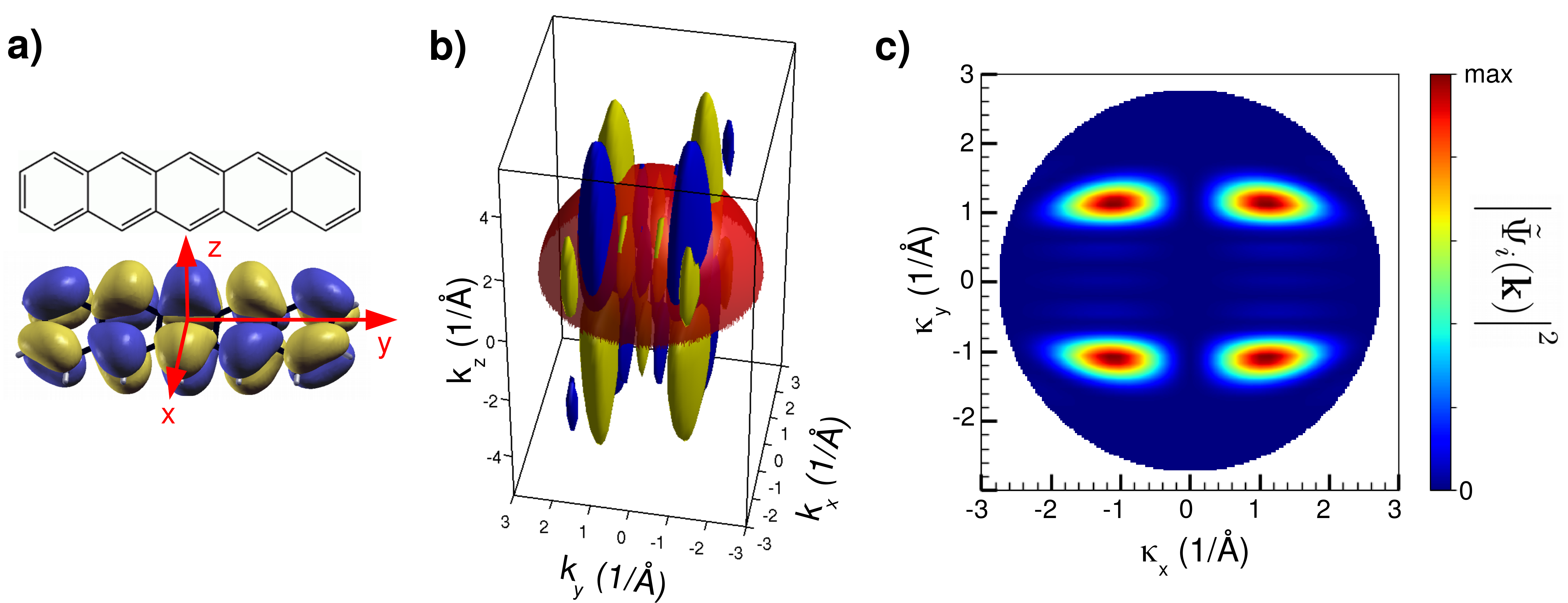}
\caption[Pentacene Maps]
{(a) Chemical structure of pentacene and its highest molecular orbital (HOMO) calculated from density-functional theory. (b) Three-dimensional Fourier transform of the pentacene HOMO orbital, yellow (blue) showing an isosurface with a constant positive (negative) value. The red hemisphere illustrates a region of constant kinetic energy (30 eV) as explained in the text. (c) Absolute value of the pentacene HOMO Fourier transform on the hemisphere indicated in panel (b). }
\label{fig:pentacene1}
\end{figure}

Figure~\ref{fig:pentacene1} illustrates the calculation of the first term for the case of the highest occupied molecular orbital (HOMO)   of pentacene (C$_{22}$H$_{14}$) \cite{Puschnig2009a}. Panel (a) depicts an isosurface representation of the HOMO  $\Psi (x,y,z)$ in real space, which is provided on a regular three-dimensional grid. Typically, some space, here $4$~{\AA}, is added in each Cartesian direction in order to safely encompass also the tails of the molecular orbital. In this example a box with a size of $L_x \times L_y \times L_z$  $\approx 12.9959 \times 22.1318 \times 8.0$~{\AA}$^3$ and a grid spacing of $\Delta x \approx \Delta y \approx \Delta z \approx 0.2$~{\AA} has been chosen, which leads to a three-dimensional array of dimensions $n_x \times n_y \times n_z = 65 \times 111 \times 41$.

The next step involves a discrete three-dimensional Fourier transform for which the function \verb|numpy.fft.fftn| from the NumPy package for numeric calculations is used \cite{Harris2020}. A subsequent shift of the zero-frequency component to the center of the array using \verb|numpy.fft.fftshift| results in the $\tilde{\Psi} (k_x,k_y,k_z)$ which is depicted in Figure~\ref{fig:pentacene1}b. Note that in this isosurface representation, the yellow and blue surfaces depict constant positive and negative values, respectively, of the real part of $\tilde{\Psi} (k_x,k_y,k_z)$. The respective grids in momentum space are given by the following definitions:
\begin{eqnarray}
\Delta k_x & = & \frac{2 \pi}{L_x}, \, k_x =  s_x+ \left( -n_x +1,-n_x+3, \cdots, n_x-3, n_x-1 \right) \frac{\Delta k_x}{2} \\
\Delta k_y & = & \frac{2 \pi}{L_y}, \, k_y =  s_y+ \left( -n_y +1,-n_y+3, \cdots, n_y-3, n_y-1 \right) \frac{\Delta k_y}{2} \\
\Delta k_z & = & \frac{2 \pi}{L_z}, \, k_z =  s_z+ \left( -n_z +1,-n_z+3, \cdots, n_z-3, n_z-1 \right) \frac{\Delta k_z}{2}
\end{eqnarray}
Note that $s_x$, $s_y$ and $s_z$ are additional shifts of half the grid spacing depending on whether the number of grid points is an even or an odd number, thus
\begin{eqnarray}
s_x & = & \left[ \mathrm{mod}(n_x,2) - 1 \right] \frac{\Delta k_x}{2} \\
s_y & = & \left[ \mathrm{mod}(n_y,2) - 1 \right] \frac{\Delta k_y}{2} \\
s_z & = & \left[ \mathrm{mod}(n_z,2) - 1 \right] \frac{\Delta k_z}{2}.
\end{eqnarray}
It is important to note that in order to increase the resolution in $k$-space, the number of grid points can be conveniently increased by zero-padding the real-space array $\Psi (x,y,z)$ during the discrete Fourier transform. Thereby, one enlarges the effective box size $L_x \times L_y \times L_z$ such that the resulting resolution in momentum space, that is $\Delta k_x,\Delta k_y$ and $\Delta k_z$, is typically in the order of $\approx 0.15$~{\AA}$^{-1}$.

In order to obtain the desired momentum map $I(k_x,k_y)$ shown in Fig.~\ref{fig:pentacene1}c for the pentacene HOMO, a hemispherical cut through the three-dimensional discrete Fourier transform $\tilde{\Psi} (k_x,k_y,k_z)$, as illustrated in Fig.~\ref{fig:pentacene1}b by the red surface, has to be computed. Selecting a final state kinetic energy $E_\mathrm{kin}$ fixes the radius of this hemisphere by the following relation
\begin{equation}
\frac{2m}{\hbar^2} E_\mathrm{kin} = k^2 = k_x^2 + k_y^2 + k_z^2.
\end{equation}
For interpolating the data on this hemisphere, a regular two-dimensional grid, denoted as $\kappa_x$ and $\kappa_y$ with a default grid spacing of $\Delta \kappa = 0.03$~{\AA}$^{-1}$ is set up and the corresponding $z$-component $\kappa_z$ is determined according to
\begin{equation}
\label{eq:kappagrid}
\kappa_z = \sqrt{\frac{2m}{\hbar^2} E_\mathrm{kin} - \kappa_x^2 - \kappa_y^2}.
\end{equation}
Then, the \verb|RegularGridInterpolator| from the SciPy Python package \cite{Jones2001} is used to compute the interpolation leading to the two-dimensional array $\tilde{\Psi} (\kappa_x,\kappa_y,\kappa_z(\kappa_x,\kappa_y))$ depicted in Fig.~\ref{fig:pentacene1}c. Note that only the absolute value of this complex-valued array $\tilde{\Psi}$ needs to be interpolated. The result of this interpolation is exemplified in Fig.~\ref{fig:pentacene1}c.

The program also allows one to vary the orientation of the molecule with respect to the coordinate frame $(x,y,z)$ introduced in Fig.~\ref{fig:arpes}. To this end, we use the three Euler angles $\varphi$, $\vartheta$ and $\psi$ and the corresponding rotation matrix
\begin{equation}
\hat{R} = \left( \begin{array}{ccc}
 \cos \varphi \cos \psi - \sin \varphi \cos \vartheta \sin \psi &  
 \sin \varphi \cos \psi + \cos \varphi \cos \vartheta \sin \psi & 
 \sin \vartheta \sin \psi\\
-\cos \varphi \sin \psi - \sin \varphi \cos \vartheta \cos \psi &  
-\sin \varphi \sin \psi + \cos \varphi \cos \vartheta \cos \psi & 
 \sin \vartheta \cos \psi\\
 \sin \varphi \sin \vartheta  &  
-\cos \varphi \sin \vartheta  & 
 \cos \vartheta 
\end{array} \right) \label{eq:Euler},
\end{equation}
\begin{figure}[tb]
\begin{center}
\includegraphics[width=\textwidth]{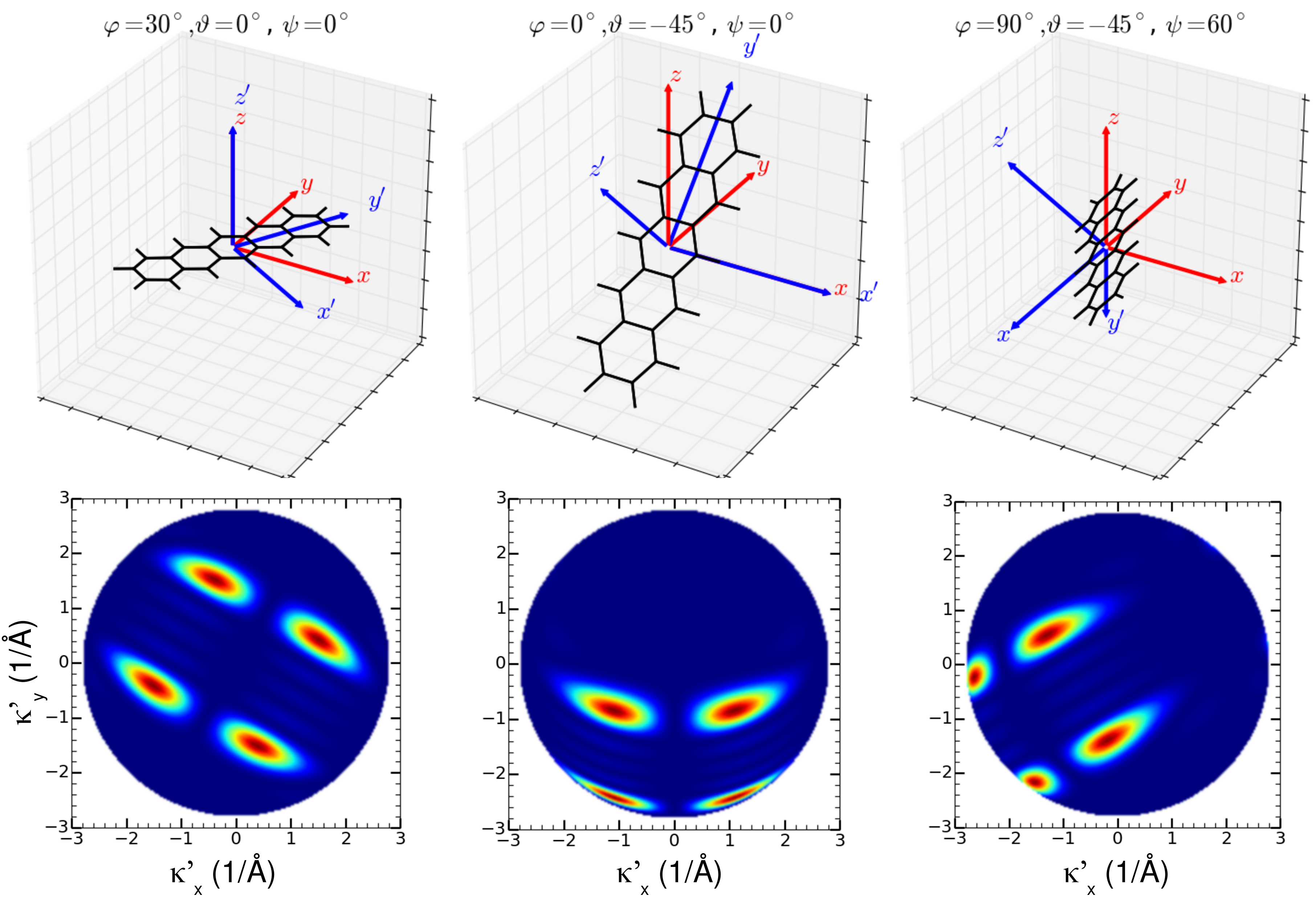}
\end{center}
\caption[Molecular orientation]
{Influence of molecular orientation on the appearance of the momentum maps. The upper row shows the effect of the three Euler angles  $\varphi$, $\vartheta$ and $\psi$ on the orientation of the molecule exemplified for pentacene for three different sets of Euler angles. Here, the red arrows indicate the fixed coordinate frame $(x,y,z)$ while the rotated one $(x',y',z')$ is depicted in blue. The lower row shows the corresponding momentum maps. }
\label{fig:orientation}
\end{figure}
In the above convention, (i) $\varphi$ represents a rotation around the $z$-axis, then (ii) $\vartheta$ represents a rotation about the new $x'$-axis and finally (iii) $\psi$ represents a rotation around the current $z'$-axis. Figure \ref{fig:orientation} illustrates the effect of these three rotations for various combinations of $\varphi$, $\vartheta$ and $\psi$.

In order to obtain the momentum maps for the rotated molecule, rather than rotating the full three-dimensional orbital either in real or momentum space, we simply rotate the hemisphere used to cut through the three-dimensional Fourier transform $\tilde{\Psi} (k_x,k_y,k_z)$  (compare Fig.~\ref{fig:pentacene1}b) by applying the transformation 
\begin{equation}
\left( \begin{array}{c} \kappa_x' \\ \kappa_y' \\ \kappa_z' \end{array} \right) = \hat{R}^\mathrm{T}\cdot 
\left( \begin{array}{c} \kappa_x  \\ \kappa_y  \\ \kappa_z  \end{array} \right).
\end{equation}
Then, the interpolation to the primed grid is executed by utilizing the \\
\verb|RegularGridInterpolator| as described above. Figure~\ref{fig:orientation} illustrates this procedure for the HOMO of pentacene for three sets of Euler angles $\varphi$, $\vartheta$ and $\psi$.

\subsubsection{Polarization factor}

As a result of the plane-wave approximation for the final state, the effect of the polarization $\ve{A}$ of the incoming photon field can be accounted for by a polarization factor $P(\kappa_x,\kappa_y)$ given by (compare Eq.~\ref{eq:PE1})
\begin{equation}
P(\kappa_x,\kappa_y) = \left|A_x \kappa_x + A_y \kappa_y + A_z \kappa_z(\kappa_x,\kappa_y) \right|^2.
\end{equation}
Here, $A_x$, $A_y$ and $A_z$ are the components of the polarization vector, which is identical to the direction of the incoming photon's electric field vector, and $\kappa_x$, $\kappa_y$ and $\kappa_z$ are the components of the electron's final state momentum vector evaluated on an identical grid as in Eq.~\ref{eq:kappagrid}. When defining the direction of the incident photon field by the two angles $\alpha$ and $\beta$, then, depending on the type of polarization, either in-plane polarization ($p$) or out-of-plane-polarization ($s$) as illustrated in Fig.~\ref{fig:arpes}, the following expressions can be derived for $P$:
\begin{figure}[tb]
\includegraphics[width=\textwidth]{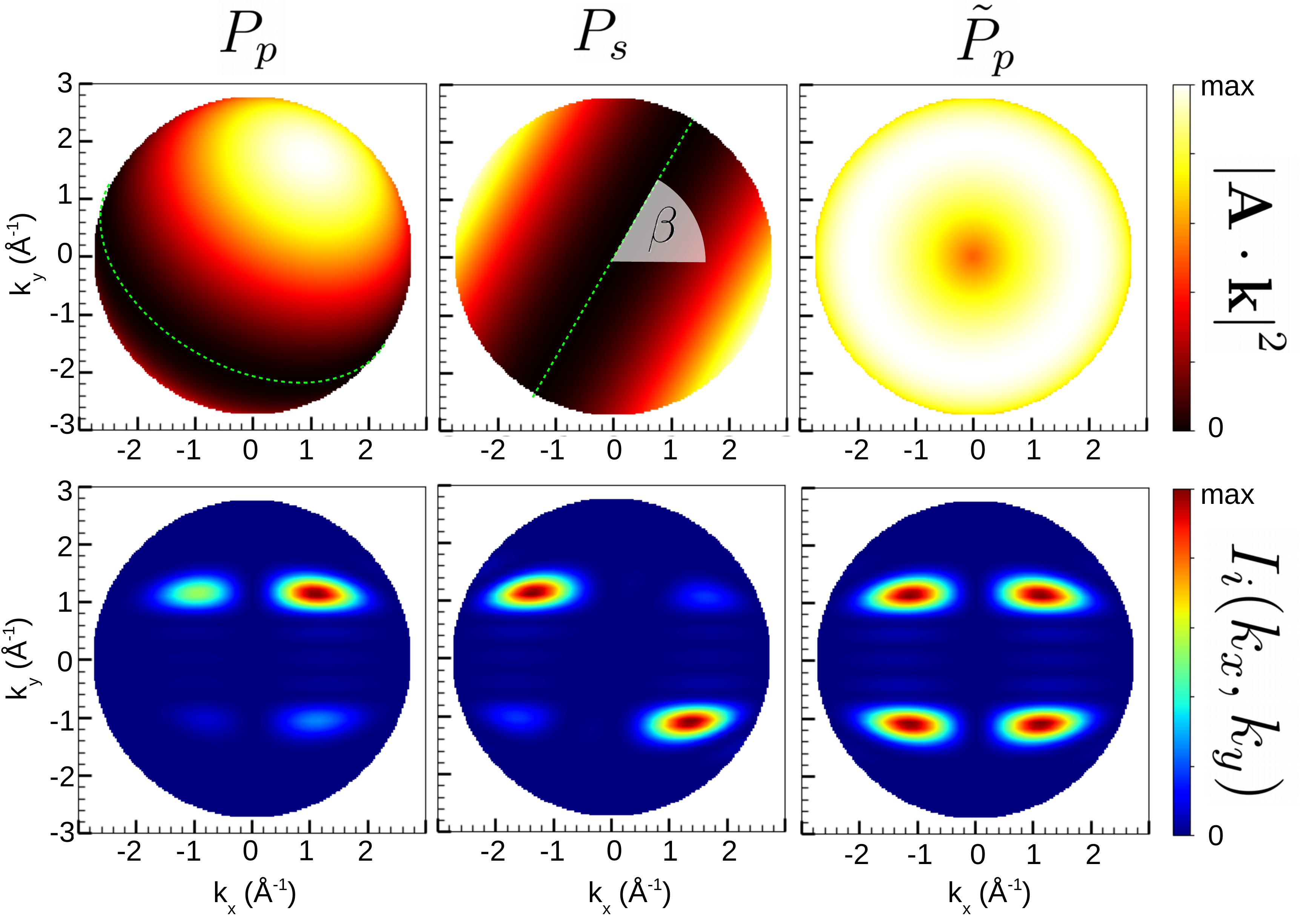}
\caption[Polarization factor]
{The effect of the polarization factor $|\ve{A}\cdot \ve{k}|^2$ on the momentum maps for an angle of incidence $\alpha=45^\circ$ and an incidence plane with $\beta=60^\circ$. The upper row illustrates this polarization factor for in-plane polarization ($P_p$), out-of-plane polarization ($P_s$) and for in-plane-polarization in the geometry the toroidal electron energy analyzer ($\tilde{P}_p$). Here, the green dashed lines indicate positions where $|\ve{A}\cdot \ve{k}|^2$ vanishes owing to  $\ve{A} \perp \ve{k}$. The lower row shows corresponding full momentum maps for the HOMO of pentacene as resulting from Eq.~\ref{eq:PE1}. }
\label{fig:polfact}
\end{figure}
\begin{eqnarray}
P_p & = & \left| \kappa_x \cos \alpha \cos \beta + \kappa_y \cos \alpha \sin \beta + \kappa_z \sin \alpha \right|^2 \label{eq:Pp}\\
P_s & = & \left|-\kappa_x \sin \beta + \kappa_y \cos \beta \right|^2 \label{eq:Ps}
\end{eqnarray}
If experimental momentum maps are measured in a geometry where the emitted electrons are collected always in the plane of incidence and the full azimuthal angular dependence of the photocurrent is obtained by rotating the sample around the substrate normal, e.g. when using the toroidal electron energy analyzer \cite{Broekman2005}, then the corresponding polarization factor simplifies to
\begin{equation}
\tilde{P}_p = \left|\sqrt{\kappa_x^2 + \kappa_y^2} \cos \alpha + \kappa_z \sin \alpha   \right|^2. \label{eq:Ptoroid}
\end{equation}

One obvious limitation of the plane-wave final state approximation is that it results in identical angular distributions of the emitted electrons for excitation with left-handed and right-handed circularly polarized light, respectively. In contrast, such a circular dichroism in the angular distribution (CDAD) has been observed experimentally and accounted for theoretically by a more sophisticated and computationally much more demanding approach which requires no assumption for the final state \cite{Dauth2016a}. However, as suggested in Ref.~\cite{Moser2016a} a simplistic approach to introduce a handedness, and therefore a CDAD signal, is to introduce an empirical damping factor, $e^{\gamma (z-z_0)}$, in the final state  which mimics the inelastic mean free path $\lambda = \frac{1}{\gamma}$ of the emitted electron \cite{Lueftner2017}. In this case, the polarization factor for right-handed ($C_{+}$) and left-handed ($C_{-}$) circularly polarized light is given by
\begin{eqnarray}
P_{C_{+}} & = & \frac{1}{2}\left(P_p + \gamma^2 \sin^2 \alpha \right) + \frac{1}{2}P_s + (\kappa_x \sin \beta - \kappa_y \cos \beta) \gamma \sin \alpha \\
P_{C_{-}} & = & \frac{1}{2}\left(P_p + \gamma^2 \sin^2 \alpha \right) + \frac{1}{2}P_s - (\kappa_x \sin \beta - \kappa_y \cos \beta) \gamma \sin \alpha 
\end{eqnarray}
Thus, the CDAD signal which is proportional to $P_{C_{+}} - P_{C_{-}}$ becomes \cite{Moser2016a}
\begin{equation}
P_\mathrm{CDAD} = +2 (\kappa_x \sin \beta - \kappa_y \cos \beta) \gamma \sin \alpha.
\end{equation}
Note that obviously $P_\mathrm{CDAD}$ vanishes if $\gamma = 0$, that is, when the inelastic mean free path $\lambda \rightarrow \infty$. By default, the inelastic mean free path is calculated from the ``universal curve'' as follows
\begin{equation}
\lambda = \frac{c_1}{E_\mathrm{kin}^2} + c_2 \sqrt{E_\mathrm{kin}}, \qquad c_1 = 1430, \, c_2 = 0.54.
\end{equation}
In this empirical formula \cite{Seah1979}, $E_\mathrm{kin}$  must be inserted in [eV] and $\lambda$ is given in [{\AA}].

\subsection{Orbital deconvolution}
\label{sec:deconvolution}

One interesting application of photoemission tomography has become known as \emph{orbital deconvolution} \cite{Puschnig2011,Willenbockel2014}. Here the energy and momentum dependence of ARPES data is utilized to deconvolute experimental spectra into individual orbital contributions. This can provide an orbital-by-orbital characterization of large adsorbate systems,  allowing one to directly estimate the effects of bonding on individual orbitals. 

The idea is quite simple and involves a least squares minimization procedure. The experimental ARPES data can be viewed as a data cube, $I_\mathrm{exp}(k_x,k_y,E_b)$, that is, the photoemission intensity is measured as a function of the two momentum components parallel to the surface, $k_x$ and $k_y$, and the binding energy $E_b$. The deconvolution of the experimental data cube then consists of minimizing the squared differences between the experimental and simulated momentum maps, 
\begin{equation}
\label{eq:chi2a}
\chi^2(w_1,w_2,\cdots, w_n) = \sum_{k_x,k_y} \left[ I_\mathrm{exp}(k_x,k_y,E_b) - \sum_{i=1}^n w_i(E_b) I_i(k_x,k_y) \right]^2 
\end{equation}
by adjusting the $n$ weights $w_i$ of all orbitals $i$ with the simulated momentum maps $I_i(k_x,k_y)$ that are allowed to contribute to the measurement data. Since the minimization is performed for each binding energy $E_b$ separately, one thereby obtains an orbital projected density of states given by the weight functions $w_i(E_b)$. Example data and the corresponding analysis will be presented in Sec.~\ref{sec:applications}.

In a similar way, a least squares minimization can be used to determine the orientation of the molecule by adjusting the Euler angles  $\varphi$, $\vartheta$ and $\psi$. Thus, one demands the sum of squared differences between an experimental momentum map and a simulated momentum map to be minimized with respect to the orientation of the molecule:
\begin{equation}
\label{eq:chi2b}
\chi^2(\varphi,\vartheta,\psi) = \sum_{k_x,k_y}  \left[ I_\mathrm{exp}(k_x,k_y,E_b) - w \, I(k_x,k_y;\varphi,\vartheta,\psi) \right]^2 
\end{equation}
Example data and the corresponding analysis will also be presented in Sec.~\ref{sec:applications}.
 
For the optimization problems defined defined in Eqs.~\ref{eq:chi2a}  and \ref{eq:chi2b}, we make use of the Python package \verb|LMFIT-py|  \cite{LMFIT}, which allows for a quite flexible implementation as demonstrated in Sec.~\ref{sec:applications}.

\subsection{Computation of molecular orbitals}
\label{sec:database}

As mentioned in Sec.~\ref{sec:initialstate}, the initial state $\Psi_i(x,y,z)$ is commonly obtained from a DFT calculation. As an example, we provide below an input file for the computation of the HOMO orbital of pentacene  when employing the open source high-performance computational chemistry code NWChem \cite{Valiev2010}:

\begin{lstlisting}[basicstyle=\footnotesize,numbers=left,numberstyle=\tiny\color{mygray}]
charge 0
geometry nocenter
  C  1.40861257   0.0         0.0
  C -1.40861257   0.0         0.0
  C  1.40791076  -2.46779313  0.0
  C -1.40791076  -2.46779313  0.0
  C  1.40791076   2.46779313  0.0
  C -1.40791076   2.46779313  0.0
  C  1.41043423  -4.94193362  0.0
  C -1.41043423  -4.94193362  0.0
  C  1.41043423   4.94193362  0.0
  C -1.41043423   4.94193362  0.0
  C  0.7286844   -1.22661113  0.0
  C -0.7286844   -1.22661113  0.0
  C  0.7286844    1.22661113  0.0
  C -0.7286844    1.22661113  0.0
  C  0.72768117  -3.67894381  0.0
  C -0.72768117  -3.67894381  0.0
  C  0.72768117   3.67894381  0.0
  C -0.72768117   3.67894381  0.0
  C  0.71646676  -6.11814601  0.0
  C -0.71646676  -6.11814601  0.0
  C  0.71646676   6.11814601  0.0
  C -0.71646676   6.11814601  0.0
  H  2.49673472   0.0         0.0
  H -2.49673472   0.0         0.0
  H  2.49613711  -2.46811675  0.0
  H -2.49613711  -2.46811675  0.0
  H  2.49613711   2.46811675  0.0
  H -2.49613711   2.46811675  0.0
  H  2.49794688  -4.9403758   0.0
  H -2.49794688  -4.9403758   0.0
  H  2.49794688   4.9403758   0.0
  H -2.49794688   4.9403758   0.0
  H  1.2479695   -7.06588298  0.0
  H -1.2479695   -7.06588298  0.0
  H  1.2479695    7.06588298  0.0
  H -1.2479695    7.06588298  0.0
end
start
basis
  * library 6-31G*
end

dft
  xc B3LYP
  convergence  nolevelshifting
  mult 1
  maxiter 2000
end

task dft energy

memory total 1000 mb 

DPLOT
TITLE pentacene_MO_73
GAUSSIAN
LimitXYZ
 -6.49794688  6.49794688   64
-11.06588298 11.06588298  110
 -4.0         4.0          40
spin total 
orbitals view; 1; 73
output 5A_MO_73
END

task dplot
\end{lstlisting}

Here, lines 3--38 define the geometry of the molecule, lines 41--43 the basis set, and lines 45--52 the exchange-correlation functional and further settings for the self-consistency cycle. The \verb|DPLOT| block in lines 56--68 results in the output of the Kohn-Sham orbital with the number 73, which is the HOMO of pentacene, on a three-dimensional real-space grid with $64 \times 110 \times 40$ intervals for a rectangular box with limits $[-6.49794688, 6.49794688]$, $[-11.06588298, 11.06588298]$ and $[-4.0,4.0]$ in $x$, $y$ and $z$-direction, respectively. Note that all lengths in this input file are give in [\AA].

The resulting output file for the HOMO orbital of pentacene is a so-called cube-file which is listed below:

\begin{lstlisting}[basicstyle=\footnotesize,numbers=left,numberstyle=\tiny\color{mygray}]
 Cube file generated by NWChem
 pentacene_MO_73
   36  -12.279342  -20.911492   -7.558906
   65    0.383729    0.000000    0.000000
  111    0.000000    0.380209    0.000000
   41    0.000000    0.000000    0.377945
    6    6.000000    2.661892    0.000000    0.000000
    6    6.000000   -2.661892    0.000000    0.000000
    6    6.000000    2.660566   -4.663453    0.000000
    6    6.000000   -2.660566   -4.663453    0.000000
    6    6.000000    2.660566    4.663453    0.000000
    6    6.000000   -2.660566    4.663453    0.000000
    6    6.000000    2.665334   -9.338900    0.000000
    6    6.000000   -2.665334   -9.338900    0.000000
    6    6.000000    2.665334    9.338900    0.000000
    6    6.000000   -2.665334    9.338900    0.000000
    6    6.000000    1.377014   -2.317959    0.000000
    6    6.000000   -1.377014   -2.317959    0.000000
    6    6.000000    1.377014    2.317959    0.000000
    6    6.000000   -1.377014    2.317959    0.000000
    6    6.000000    1.375118   -6.952196    0.000000
    6    6.000000   -1.375118   -6.952196    0.000000
    6    6.000000    1.375118    6.952196    0.000000
    6    6.000000   -1.375118    6.952196    0.000000
    6    6.000000    1.353926  -11.561620    0.000000
    6    6.000000   -1.353926  -11.561620    0.000000
    6    6.000000    1.353926   11.561620    0.000000
    6    6.000000   -1.353926   11.561620    0.000000
    1    1.000000    4.718144    0.000000    0.000000
    1    1.000000   -4.718144    0.000000    0.000000
    1    1.000000    4.717015   -4.664064    0.000000
    1    1.000000   -4.717015   -4.664064    0.000000
    1    1.000000    4.717015    4.664064    0.000000
    1    1.000000   -4.717015    4.664064    0.000000
    1    1.000000    4.720435   -9.335957    0.000000
    1    1.000000   -4.720435   -9.335957    0.000000
    1    1.000000    4.720435    9.335957    0.000000
    1    1.000000   -4.720435    9.335957    0.000000
    1    1.000000    2.358320  -13.352583    0.000000
    1    1.000000   -2.358320  -13.352583    0.000000
    1    1.000000    2.358320   13.352583    0.000000
    1    1.000000   -2.358320   13.352583    0.000000
  0.00000E+00  0.00000E+00  0.00000E+00  0.00000E+00  0.00000E+00  0.00000E+00
  0.00000E+00  0.00000E+00  0.00000E+00  0.00000E+00  0.00000E+00  0.00000E+00
  ...
\end{lstlisting}
Here, lines 3--42 define the box size, real space grid and the molecular structure where length units are now atomic units, that is Bohr. The actual data points of the real-valued three dimensional array for $\Psi_i(x,y,z)$ on a regularly spaced $65 \times 11 \times 41$ grid starts in lines 43 and continues until the end of the file (not shown).

It should be noted that, as a related effort but independent of \verb|kMap.py|, the authors have also set up a database of molecular orbitals for prototypical organic $\pi$-conjugated molecules which is based on the atomic simulation environment (ASE) \cite{Larsen2017} with NWChem as a calculator. This ``Organic Molecule Database'' \cite{Puschnig2020} can be accessed via a web-interface from which cube files containing the necessary input for \verb|kMap.py| can be downloaded, but there exists also a direct link in \verb|kMap.py| allowing for a comfortable import of molecular orbitals from this online database.

\section{Applications}
\label{sec:applications}

\subsection{Pentacene tilt angle in a multilayer film}
\label{sec:tiltangle}
\begin{figure}[tb]
\includegraphics[width=\textwidth]{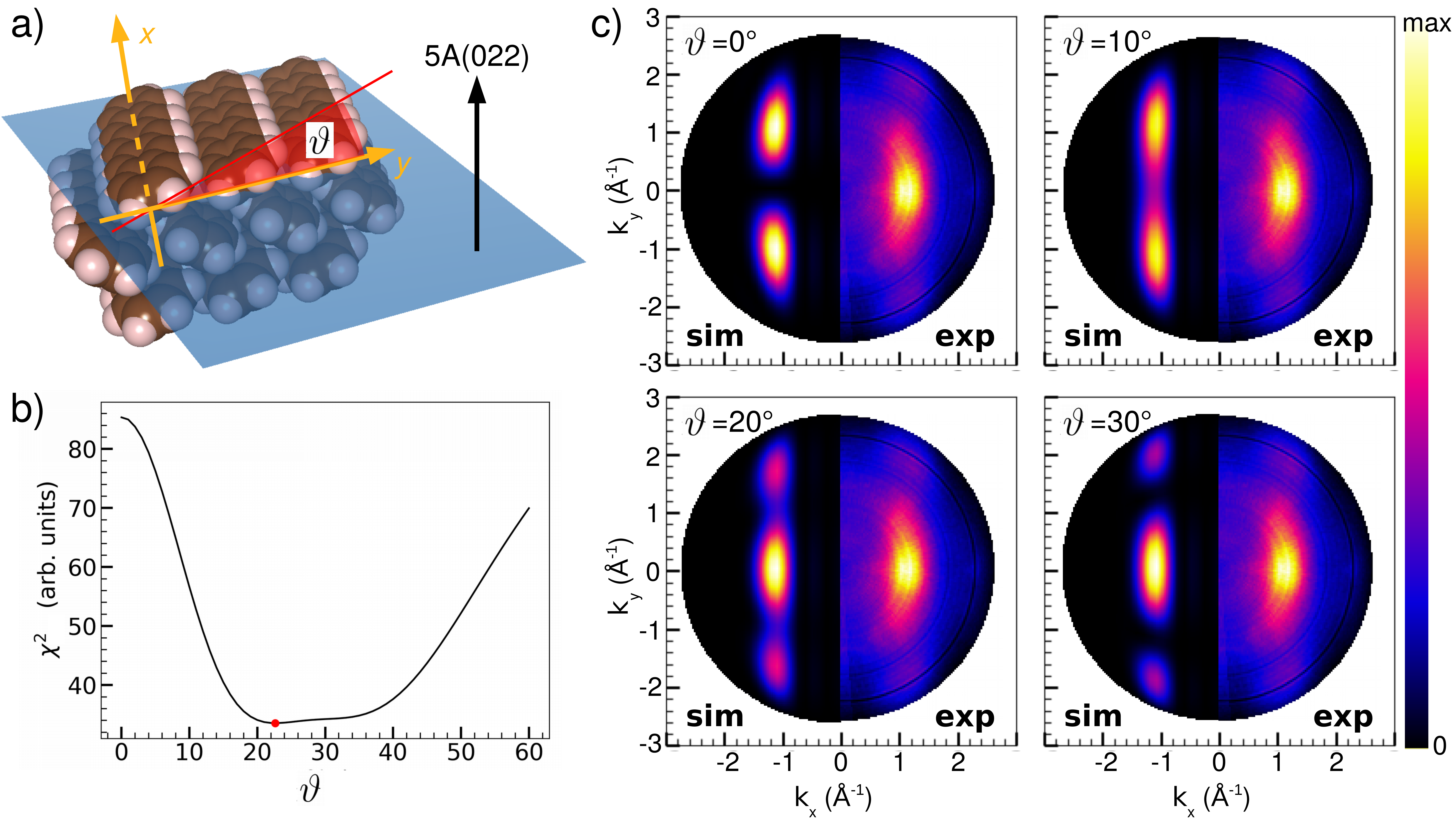}
\caption[Pentacene Maps]
{(a) Geometry of the (022) plane of the pentacene crystal structure exhibiting molecules oriented with their long moelcular axis parallel to the surface plane and with a tilt angle of $\pm \vartheta$. (b) Sum of least squares $\chi^2$ according to Eq.~\ref{eq:chi2b} as a function of the molecular tilt angle $\vartheta$. The best fir for $\vartheta$ is indicated by the red dot. (c) Comparison of experimental (right half) and  simulated momentum maps (right half) for four exemplary tilt angles $\vartheta=0^\circ,10^\circ,20^\circ$ and $30^\circ$. Note that the simulated momentum maps are calculated as superpositions of $+\vartheta$ and $-\vartheta$.}
\label{fig:pentacenetilt}
\end{figure}

In this subsection we illustrate how the minimization of Eq.~\ref{eq:chi2b} can be used to determine the molecular orientation in molecular films by photoemission tomography \cite{Puschnig2009a}. Specifically, we consider a pentacene multilayer which is formed when the molecule is vacuum deposited on a p$(2 \times 1)$ oxygen reconstructed Cu(110) surface. Thereby its long axis orients parallel to the oxygen rows, resulting in crystalline pentacene(022) films \cite{Koini2008} as illustrated in Fig.~\ref{fig:pentacenetilt}a. In this film, the molecular planes are tilted out of the surface plane by an angle of $\pm \vartheta$. 
 
The momentum map of the HOMO of pentacene, already introduced in Fig.~\ref{fig:pentacene1}, can indeed be used to determine the tilt angle $\vartheta$ of the molecule. As can be seen from Fig.~\ref{fig:pentacenetilt}b and \ref{fig:pentacenetilt}c, the simulated momentum maps are quite sensitive to this tilt. Already a visual comparison of the simulated maps for various angles of $\vartheta=0^\circ,10^\circ,20^\circ$ and $30^\circ$ with the experimental data (Fig.~\ref{fig:pentacenetilt}c) reveals that the best agreement can be found for a tilt angle somewhere between $20^\circ$ and $30^\circ$. Indeed, the minimization of $\chi^2$ according to Eq.~\ref{eq:chi2b} leads to an optimal value of $\vartheta_\mathrm{opt}=22.7^\circ$. Note that in the simulated maps superpositions of tilts for $+\vartheta$ and $-\vartheta$ have been considered to account for the two-fold symmetry of the measured momentum maps, and that in the minimization of $\chi^2$ also a $k$-independent constant has been included in the optimization to respect background emissions in the experimental data.
The so-obtained value is in good agreement with a value of $\vartheta_\mathrm{xray}=26^\circ$ determined from an x-ray diffraction structure solution for pentacene crystal \cite{Koini2008}. The present value also agrees with a previously determined one from photoemission tomography \cite{Puschnig2009a}, where only linescans through momentum maps but not the full two-dimensional momentum maps have been used in the minimization of $\chi^2$.

\subsection{Orbital deconvolution of the M3-emission of PTCDA/Ag(110)}
\label{sec:M3}
\begin{figure}[tb]
\includegraphics[width=\textwidth]{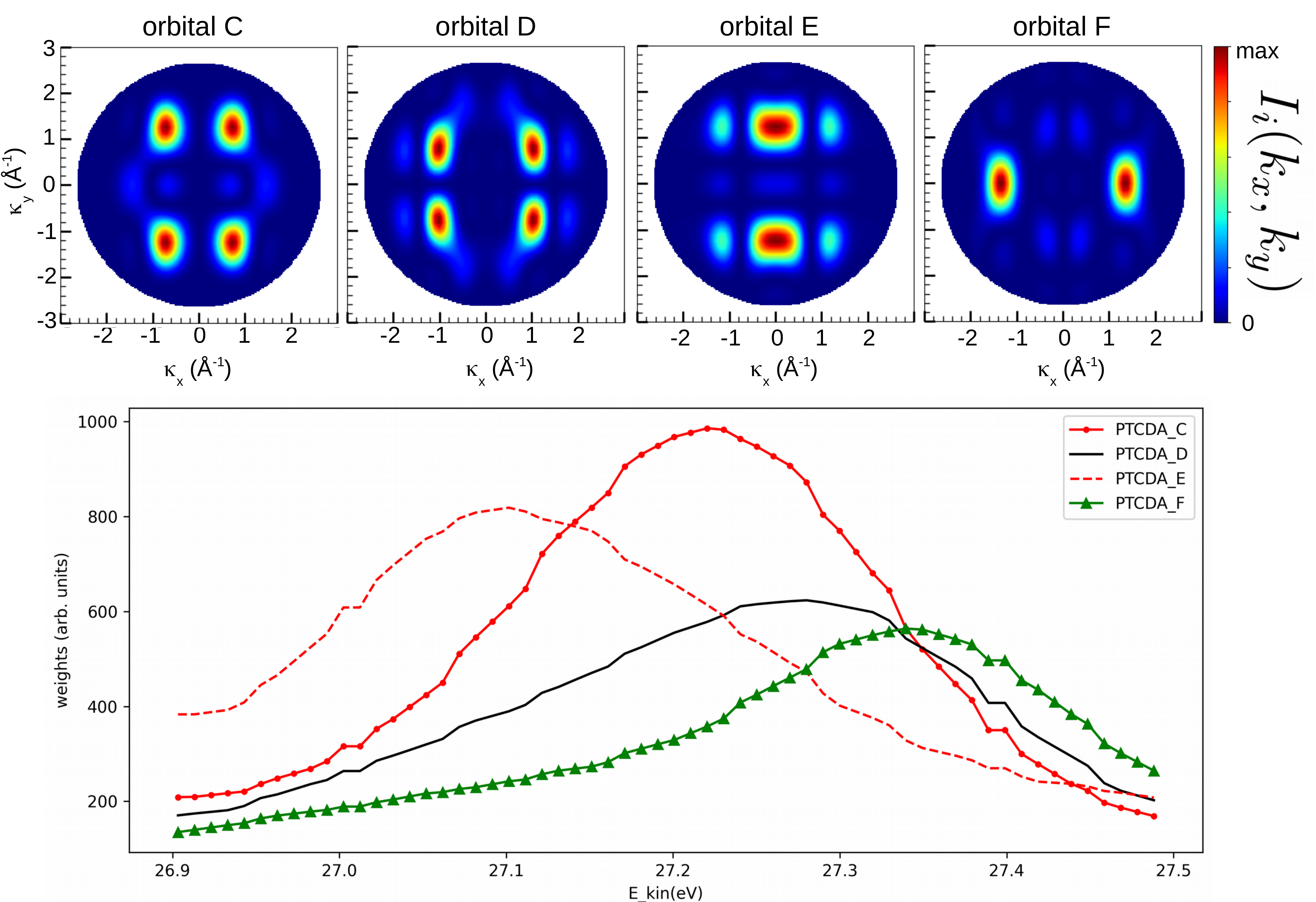}
\caption[Orbital deconvolution]
{Orbital deconvolution of the M3-emission of PTCDA/Ag(110). The top row shows the momentum maps of four PTCDA's orbitals denoted as C, D, E and F. Minimizing Eq.~\ref{eq:chi2a} separately for each kinetic energy $E_\mathrm{kin}$ leads to the orbital weights $w_C$, $w_D$, $w_E$ and $w_F$ plotted in the lower part of the figure.}
\label{fig:deconvolution}
\end{figure}

In this subsection, we demonstrate the capabilities of \verb|kMap.py| to perform an orbital deconvolution as described in Sec.~\ref{sec:deconvolution}. To this end, we use the prototypical case of the molecule PTCDA (3,4,9,10-perylene-tetracarboxylic-dianhydride) adsorbed on a Ag(110) \cite{Puschnig2011}. For a monolayer of PTCDA/Ag(110), photoemission experiments have revealed a molecular emission centered around an energy of $E_b \approx -3.4$~eV below the Fermi energy, which has been termed the ``M3 feature'' in Ref.~\cite{Puschnig2011}. The M3-emission has been shown to originate from four molecular orbitals which are closely spaced in energy, denoted as orbitals C, D, E and F. Using the prescription described in Sec.~\ref{sec:method}, the momentum maps for these four orbitals can be calculated based on isolated molecule DFT results for PTCDA \cite{Puschnig2020}. Using photons incident with $\alpha=40^\circ$, the polarization factor $\tilde{P}_p$ suitable for the toroidal electron energy analyzer geometry, and  a final state kinetic energy of the emitted electron $E_\mathrm{kin}=27.2$~eV, the four momentum maps depicted in Fig.~\ref{fig:deconvolution} have been obtained.

The experimental data set, that is, the photoemission intensity $I_\mathrm{exp}(k_x,k_y,E_\mathrm{kin})$ as a function of the two parallel momentum components $k_x$ and $k_y$ and the kinetic energy $E_\mathrm{kin}$ of the emitted electron, is provided over an  energy range from about $26.9$ to $27.5$~eV with a spacing of $\Delta E_\mathrm{kin} = 0.025$~eV. For each kinetic energy slice, first the experimental data $I_\mathrm{exp}(k_x,k_y,E_\mathrm{kin}=\mathrm{const})$ and the simulated momentum maps $I_i(k_x,k_y)$ are interpolated to the same $k$-grid using the \verb|RegularGridInterpolator|. Then the minimization of Eq.~\ref{eq:chi2a} is performed with \verb|LMFIT-py| \cite{LMFIT} leading to the four weights $w_C$, $w_D$, $w_E$ and $w_F$ plotted in the lower part of Figure~\ref{fig:deconvolution}. In order to account for a background in the experimental data, a simple, $k$-independent background  has been included in the optimization procedure. The resulting dependence $w_i(E_b)$ can be interpreted as an orbital-projected density of states and is known as the orbital deconvolution which has already been described for the M3-feature of PTCDA/Ag(110) in Ref.~\cite{Puschnig2011}.

\section{Conclusion and outlook}
\label{sec:outlook}

Arguably, the most direct method of addressing the electronic properties in general and the density of states in particular is ultra-violet photoemission spectroscopy. Recent applications of angle-resolved  photoemission spectroscopy to large aromatic molecules have shown that the angular distribution of the photoemission intensity can be easily understood with a simple plane wave final state approximation.  Over the last decade, the method of orbital or photoemission tomography  has emerged and developed as a combined experimental/theoretical approach seeking an interpretation of the photoemission distribution in terms of the initial state wave functions. This paper presents the open source python-based program \verb|kMap.py| which comes with an easy-to-use graphical user interface in which simulated momentum maps can be directly compared with experimental ARPES data. Among various applications,  \verb|kMap.py| can be used to unambiguously assign emissions to particular molecular orbitals, and in particular, to deconvolute measured spectra into individual orbital contributions (Sec.~\ref{sec:M3}), and to extract detailed geometric information (Sec.~\ref{sec:tiltangle}). Other applications currently available in \verb|kMap.py|, but not demonstrated herein, would be the precise determination of the charge balance and transfer at the organic/iorganic interfaces \cite{Hollerer2017}. 

For the future, we envision further developments and applications as well as a growing number of potentially interested users of \verb|kMap.py|. On the experimental side, there has been significant progress in photoemission spectrometers and excitation sources which boost experimental momentum space imaging of the electronic properties. This includes spin-sensitive detectors, photoemission momentum microscopes, time-resolved photoelectron spectrometers to be combined with laser excitations sources for pump-probe experiments. The latter will open a completely new window into the photoemission from excited states above the Fermi level and their time evolution \cite{Wallauer2020}. Understanding these new experimental results will require appropriate simulations tools as provided by \verb|kMap.py|.  

Future directions for developments in \verb|kMap.py| include the reconstruction of real space orbital distributions from measured momentum map. The  necessary numerical retrieval of the phase information has been demonstrated already in a number of publications \cite{Luftner2013,Kliuiev2016a,Kliuiev2018a,Jansen2020} and should be straight-forward to implement also into \verb|kMap.py|. A more fundamental aspect of photoemission tomography concerns the underlying approximation of the final state as a plane wave which certainly has known limitations \cite{Bradshaw2015}. Here, it will be desirable to provide computationally tractable alternatives which go beyond the plane-wave approximation also within the  simulation platform \verb|kMap.py|.

\section*{Acknowledgment}
Financial support from the Austrian Science Fund (FWF) (project I3731) is gratefully acknowledged.





\section*{References}


\end{document}